\documentclass{aa}  

\usepackage{graphicx}
\usepackage{txfonts}
\usepackage{xcolor}
\usepackage{amsmath}
\usepackage{ulem} 
\usepackage[colorlinks=true,citecolor=blue,urlcolor=blue]{hyperref}
\begin{document}

\title{{\it Gaia} {\sc dr3} reveals the complex  dynamical evolution within star clusters }
\titlerunning{Spatial evolution of star clusters}
\authorrunning{Viscasillas V{\'a}zquez et al.}


 \author{
C. Viscasillas V{\'a}zquez\inst{\ref{vilnius}}, 
L. Magrini\inst{\ref{oaa}}, N. Miret-Roig\inst{\ref{vienna}}, N. J. Wright\inst{\ref{keele}}, 
J. Alves\inst{\ref{vienna}}, 
L. Spina\inst{\ref{oapd}},  R. P. Church\inst{\ref{lund}},  G. Tautvai{\v s}ien{\. e}\inst{\ref{vilnius}}, \and S. Randich\inst{\ref{oaa}} }

\institute{
Institute of Theoretical Physics and Astronomy, Vilnius University, Sauletekio av. 3, 10257 Vilnius, Lithuania. \label{vilnius}
\and 
INAF - Osservatorio Astrofisico di Arcetri, Largo E. Fermi 5, 50125, Firenze, Italy.  \label{oaa} 
\and
University of Vienna, Department of Astrophysics, Türkenschanzstraße 17, 1180 Wien, Austria.
\label{vienna}
\and
Astrophysics Group, Keele University, Keele ST5 5BG, UK.
\label{keele}
\and
Lund Observatory, Division of Astrophysics, Department of Physics, Lund University, Box 43, SE-22100 Lund, Sweden 
\label{lund}
\and
INAF - Padova Observatory, Vicolo dell'Osservatorio 5, 35122 Padova, Italy\label{oapd}
}

   \date{Received 2 May 2024 / Accepted 28 June 2024 }

 
  \abstract
   {Star clusters, composed of stars born from the same molecular cloud, serve as invaluable natural laboratories for understanding the fundamental processes governing stellar formation and evolution. }
   {This study aims to investigate correlations between the Mean Interdistance ($\bar{D_{i}}$), Mean Closest Interdistance ($\bar{D_{c}}$) and Median Weighted Central Interdistance ($\bar{D_{cc}}$) with the age of star clusters, examining their evolutionary trends and assessing the robustness of these quantities as possible age indicators.} 
   {
   We selected a sample of open clusters in the solar region and with a representative number of members (e.g. well populated and without outliers). The interdistances are derived from the spatial distribution of  member stars within a cluster.
   Their  evolution over time allows us to use them as an age indicators for star clusters.}
   {Our investigation reveals a high-significant correlation between the interdistances and cluster age. 
   Considering the full sample of clusters between 7 and 9 kpc, the relationship is very broad. This is 
   due to uncertainties in parallax, which increase with increasing distance. In particular, we must limit the sample to  a maximum distance from the Sun of about 200 pc to avoid artificial effects on cluster shape and on the spatial distribution of their stars along the line of sight.}
   {
   By conservatively restraining the distance to a maximum of $\sim$200 pc, we have established a relationship between the interdistances and the age of the clusters. In our sample, the relationship is mainly driven by the internal expansion of the clusters, and is marginally affected by external perturbative effects. 
   Such relation might enhance our comprehension of cluster dynamics and might be used to derive cluster dynamical ages.}

   \keywords{Galaxy: disc; Galaxy: evolution; Galaxy: abundances; Galaxy: kinematics and dynamics; open clusters and associations: general.}

   \maketitle
%

\section{Introduction}

Star clusters are fundamental tools in many astrophysical fields. They  offer invaluable insights on the mechanisms of stellar mass assembly in galaxies  \citep[e.g.][]{Krumholz2019ARA&A..57..227K, Adamo2020SSRv..216...69A}. Most stars are, indeed,  born in clusters \citep{Lada2003}. So their study is fundamental to our understanding of both the star formation process and the dynamical evolution from bound stellar populations to the unbound ones.  
Their formation, evolution and survival are related to both intrinsic factors and to the surrounding environment \citep[e.g.][]{Bastian2012MNRAS.419.2606B, Viscasillas23}. In the present work, we aim at understanding the role of the internal dynamics and the time evolution of the spatial distribution of members within star clusters. These are, indeed, crucial aspects to unravel the processes governing their existence and survival over time \citep[e.g][]{Grudi2023MNRAS.519.1366G, Rodriguez2023MNRAS.521..124R}. 

In this context, various metrics have been employed to characterise the internal structure of stellar clusters, each offering unique perspectives on their nature. Traditionally, the Minimum Spanning Tree (MST), introduced by \citet{Prim1957}, has served until today as a prominent tool for analysing the spatial connectivity of member stars within clusters \citep[e.g.][]{Maurya23}. The MST, a graph-theoretic construct, elucidates the shortest path connecting all member stars without forming closed loops. Based on it, the Q-parameter, introduced by \citet{Cartwright2004}, combines normalised mean edge length ($\bar{m}$) of the MST with the spatial dispersion of stars ($\bar{s}$), expressed as Q = $\bar{m}$/$\bar{s}$. This metric distinguishes between large-scale radial density gradients and multiscale (fractal) subclustering, offering a comprehensive assessment of stellar cluster structure. The MST also served as a basis for \citet{Allison2009} to introduce a novel method to detect and quantify mass segregation in star clusters by comparing it in massive stars with that of randomly selected stars. If the mass segregation is present, the MST length of the most massive stars would be shorter than that of random stars, since the massive stars are more centrally concentrated than lower mass stars. Modified versions of the MST were also presented by, e.g., \citet{Olczak2011} and \citet{Yu2011} to measure the degree of mass segregation. Thus, the first of them uses a Delaunay triangulation in two dimensions to construct a graph of stellar positions projected onto a plane. 
Besides, \citet{Parker2014} and \citet{Wright2014} demonstrated the utility of the MST in discerning between dynamically unevolved (substructured) and dynamically evolved (smooth) spatial distributions. The latter study further revealed that OB associations exhibit substructure, indicating their dynamical youth, unmixed state, and excluding them from being the expanded remnants of single star clusters.

During the initial few million years, young clusters exhibit considerable expansion, as proposed by \citet{Bastian2008}. This was also observed by \citet{Kuhn2019}, and more recently by \citet{DellaCroce2023}, taking advantage of the unprecedented amount of precise data from the {\it Gaia} catalogue. The dynamical evolution of clusters was also addressed by \citet{Angelo2021} who investigated 38 Galactic open clusters. Their study employed structural and time-related parameters associated with the clusters' dynamical evolution, such as core, tidal, and half-mass radii, ages, and crossing times. Their results suggest that dynamically older systems tend to be more centrally concentrated and are less subject to the tidal disruption. This seems to be supported by the results of \citet{Tarricq2022} using {\it Gaia} EDR3, who found that, on average, older clusters have smaller core sizes compared to younger ones. 
However, the overall size of clusters appears to slightly increase with age, while the fraction of stars in the halo decreases. \citet{Tarricq2022} emphasises that parameters like cluster sizes and mass segregation levels are age-dependent and cannot be simplified as single functions of time.

Using N-body simulations, \citet{Wilkinson2003} investigated the variations of the core radii among intermediate-age and old star clusters in the Large Magellanic Cloud. Their simulations show that clusters on circular and elliptical orbits exhibit a similar evolution of their core radius and that the tidal field of the Large Magellanic cloud (LMC) has not yet acted in the intermediate age clusters. It is worth mentioning that \citet{Pang2020} found that clusters born in the same giant molecular cloud and with similar ages can potentially have divergent futures. This suggests that despite sharing common origins, various factors such as initial dynamical states prior to gas expulsion can lead to distinct evolutionary paths for different clusters. 
 Other studies have examined the effect of orbital eccentricity on the dynamical evolution of star clusters \citep{Cai_2016,Ebrahimi_2019}. Previously, \citet{webb14} demonstrated that increasing orbital eccentricity decelerates cluster evolution due to a weaker mean tidal field at a given perigalactic distance. However, although eccentric orbits may be affected by a weaker gravitational field, the effects of perigalactic passes and tidal heating can partially offset this decrease. Considering that star clusters spend most of their lifetimes near the apogalacticon, the characteristics of clusters that appear highly dynamically evolved for a given galactocentric distance can be attributed to an eccentric orbit. On the other hand, \citet{Martinez-Medina_2017} uncovered the vulnerability of high-altitude open clusters due to severe tidal destruction upon crossing the Galactic disc. Despite their distance from in-plane substructures, clusters above 200 pc face significant tidal shocks, peaking around 600 pc before declining due to reduced encounters with the disc.

Besides, the evaporation of the clusters occurs at different dissolution rates, depending on their location. So for example, star clusters within 150 pc from the Galactic center can dissolve in approximately 50 Myr, while in the solar neighbourhood, most open clusters evaporate completely in less than 1 Gyr \citep[see e.g.][and references therein]{Pavani_2007}.
Indeed, mass segregation within clusters can manifest either as a primordial characteristic, with the formation of clusters featuring the concentration of the most massive stars at or near the center, or dynamically, as a result of post-formation migration driven by two-body interactions \citep{Bonnell2001, Allison2009}. Nonetheless, \citet{Dib2018} based on the MST and Q-parameter for a large sample of clusters, suggest that the majority of clusters neither exhibit strong concentration nor display significant substructure. Moreover, they did not find any correlation between the structure of clusters, the extent of mass segregation and their position within the Galaxy.
\citet{Pang2022} employed the  StarGO (Stars' Galactic Origin) algorithm with {\it Gaia} {\sc eDR3} data to examine morphology and kinematics of 85 open clusters, categorising substructures beyond the tidal radius into four types based on age and characteristics: filamentary  and fractal  for clusters <100 Myr, and halo and tidal tail  for clusters >100 Myr. On the other hand, the study of \citet{Hu2021} suggested that clusters with ellipticity > 0.4 have a tendency to deform or stretch in the direction of the Galactic plane.

In addition, the study of the evolution of cluster structure gives us both information on its expansion and dispersion over time, and at the same time, provides an empirical way of evaluating the age of clusters. 
There are numerous methods for measuring the ages of clusters based on their being composed of simple and therefore coeval stellar populations. 
The most widely used method is that of isochrone fitting, both using  observational quantities, such as magnitudes and colours, and derived stellar  parameters. 
However, this method is prone to uncertainties when the cluster sequence is not well populated \citep{bossini2019A&A...623A.108B, CantatGaudin20, Cavallo24}, as well as to model uncertainties, differential extinction, binarity, variability, etc. 
Methods based on the kinematics of the cluster members can therefore be supportive and helpful. 
Very recently, \citet{Miret2023} addressed the challenge of determining star ages by comparing isochronal and dynamical methods for very young clusters. The authors found a consistent difference of $\sim$5.5 Myrs, suggesting that this reflects the  time during which young stars remain bound to their birth cloud before moving away. The combination of these two methods provides a valuable tool for constraining evolutionary models, offering insights into the impact of local conditions and stellar feedback on the stellar cluster formation and dispersal. Other methods are based on the chemical properties of cluster members, as for instance the [C/N] ratio in giant stars \citep[e.g.][]{Casali2019A&A...629A..62C, Spoo2022AJ....163..229S}, the [$s$-process/$\alpha$] ratio \citep[e.g.][]{Casali2020A&A...639A.127C, Viscasillas_2022}, and the lithium abundances \citep[e.g.][]{Jeffries2023MNRAS.523..802J}. These methods need to be calibrated on clusters with known ages, usually from isochrone fitting. In addition, asteroseismology applied to both field stars and clusters is also improving our knowledge of stellar ages \citep[][]{Miglio2021ExA....51..963M, Palakkatharappil_2023}.

In the present work, we investigate the correlation between the internal spatial properties  of cluster members,  such as their distances between them, the so-called interdistances, and the age and orbital properties of clusters. 
The paper is organised as follows: Section \ref{sec:sample} introduces our initial sample and the filtering processes adopted  to obtain  our dataset. In Section \ref{sec:interdistances}, we define three different types of interdistances and we compute them for our sample.   Section $\ref{sec:distance}$ investigates the distance limits within which we can safely study the spatial structure of clusters.  After defining our benchmark cluster sample, in Section \ref{sec:interditances_vs_agewe} we investigate the  evolution of interdistances with cluster age. Section \ref{sec:orbits} studies the dependence on age of cluster orbital parameters, as eccentricity and the maximum height above the Galactic plane. Finally, Section \ref{sec:conclusions} provides a summary of our results and presents the main conclusions drawn from this study.




\section{The sample}
\label{sec:sample}
We started from the sample of $\sim$1,300,000 members of clusters in \citet{Hunt23} and we selected stars with a high probability of belonging to a given  cluster (Prob > 0.9) and located within the estimated tidal radius ('inrt' = 1). 
This reduced the sample to approximately $\sim$380,000 member stars belonging to approximately $\sim$7,000 clusters. For the given sample, and employing the {\it Gaia} {\sc DR3} distances \citep{bailerjones21}, we computed the galactocentric coordinates using {\sc Astropy} \citep{Astropy_2018}. We excluded member stars identified as outliers within clusters by applying the interquartile range (IQR) rule, in a similar way as the approach outlined by \citet{Viscasillas_2022} in the chemical space, but in this case according to their positions.  We adopted  the clusters ages and galactocentric distances of the clusters  provided by \citet[][hereafter C24]{Cavallo24}. We discarded clusters classified as globular clusters (g) and moving groups (m), according to the classification provided in \citet{Hunt23}. We kept in our sample only open clusters (o). 

To have a sample of clusters that have originated from the same region of the disc, and therefore with a similar dynamical and chemical evolution, we selected clusters located in the solar annulus, i.e., at a Galactocentric distance ranging between 7 and 9 kpc. We obtained  a sample of $\sim$3,000 clusters. We examined the impact of varying member counts across different quantile ranges and we found that the results are almost independent of the selected quantile ranges. Among these clusters, we selected those with a typical number of members, thus excluding the very populated or very small ones, and including those with a number of members falling within the interquartile range (Q1-Q3), i.e. from 22 to 72 members. By doing so, we minimised the impact of extreme values, resulting in a more balanced representation of open clusters in terms of their membership size.  Our final sample is composed of  $\sim$1,500 open clusters.
From here on, we will refer to this sample as "our sample". 
For clusters in our sample we computed the orbits with {\sc galpy} and the Galactic potential {\sc MW2014} \citep{bovy15}, using the clusters mean radial velocities, proper motions and distances from {\it Gaia} {\sc dr3}, obtained from the computation of \citet{Hunt23}. We assumed a solar position (R, Z)$_{\odot}$=(8.249, 0.0208) kpc and solar cylindrical velocity components (V$_R$, V$_{\phi}$, V$_Z$)$_{\odot}$=(9.5, 250.7, 8.56) km s$^{-1}$ \citep{gravity20}.


\section{The mean interdistances as a cluster parameter}
\label{sec:interdistances}
The degree of compactness, the distance between members and their spatial distribution give us information about the evolutionary status of star clusters. There are various quantities that can be used, as shown in the Introduction. In this paper, we present the distances between members as indicators of cluster evolution, in three different forms: mean interdistance, mean closest interdistance and median weighted central interdistance. 
Each of the three interdistances provides distinct insights into the spatial distribution and structure of the clusters. The mean interdistance ($\bar{D_{i}}$) offers a measure of the overall spacing among cluster members, capturing the average separation between any two stars within a given  cluster. The mean closest interdistance ($\bar{D_{c}}$) highlights the proximity of each member to its nearest neighbour, indicating the local density or compactness of the cluster. On the other hand, the median weighted central interdistance ($\bar{D_{cc}}$) focuses on the distance of each member to the cluster center, revealing the degree of concentration or dispersion around the central region. These metrics collectively provide a comprehensive characterization of the internal structure and arrangement of stars within the clusters.

\subsection{The mean interdistance}
We introduce  the parameter "mean interdistance" ($\bar{D_{i}}$), defined as the average spatial separation between member stars within each cluster (see Eq. \ref{eq:Di}). The mean interdistance is calculated as the average distance between all unique pairs of member stars within a given cluster in three-dimensional space. Mathematically, for a cluster with n stars and Cartesian coordinates ($x_{i}, y_{i}, z_{i}$) for each star, the interdistance  is expressed as:


\begin{align}
\label{eq:Di}
\bar{D_i} &= \frac{\sum_{i=1}^{n} \sum_{j=1,j\neq i}^{n} d_{ij}}{n(n-1)/2} \nonumber \\
&= \frac{2}{n(n-1)} \sum_{i=1}^{n} \sum_{j=1,j\neq i}^{n} \sqrt{(x_i - x_j)^2 + (y_i - y_j)^2 + (z_i - z_j)^2}
\end{align}

where $d_{ij}$ represents the distance between each unique pair of stars (i,j) within the cluster. This metric provides information on the spatial distribution  of stars within individual clusters.

\subsection{The mean closest interdistance}
We define the concept of "mean closest interdistance", denoted as $\bar{D_{c}}$, which represents the average distance of each member star within a cluster to its nearest neighbour (Eq. \ref{eq:Dc}). Mathematically, for a cluster having n stars with Cartesian coordinates ($x_{i}, y_{i}, z_{i}$), the mean closest distance is calculated as:


\begin{align}
\label{eq:Dc}
\bar{D_c} &= \frac{\sum_{i=1}^{n} \sum_{j=1,j \neq i}^{n} \min_{j \neq i} d_{c_{ij}}}{n(n-1)} \nonumber \\
&= \frac{1}{n(n-1)} \sum_{i=1}^{n} \sum_{j=1,j\neq i}^{n} \sqrt{(x_i - x_j)^2 + (y_i - y_j)^2 + (z_i - z_j)^2}
\end{align}

where $d_{c_{ij}}$ represents the distance from each star to its closest neighbour within the same cluster. This metric provides insights into the average proximity of stars within individual clusters, offering valuable information on the internal structure of these stellar groupings.

\begin{figure*} 
  \resizebox{\hsize}{!}{\includegraphics{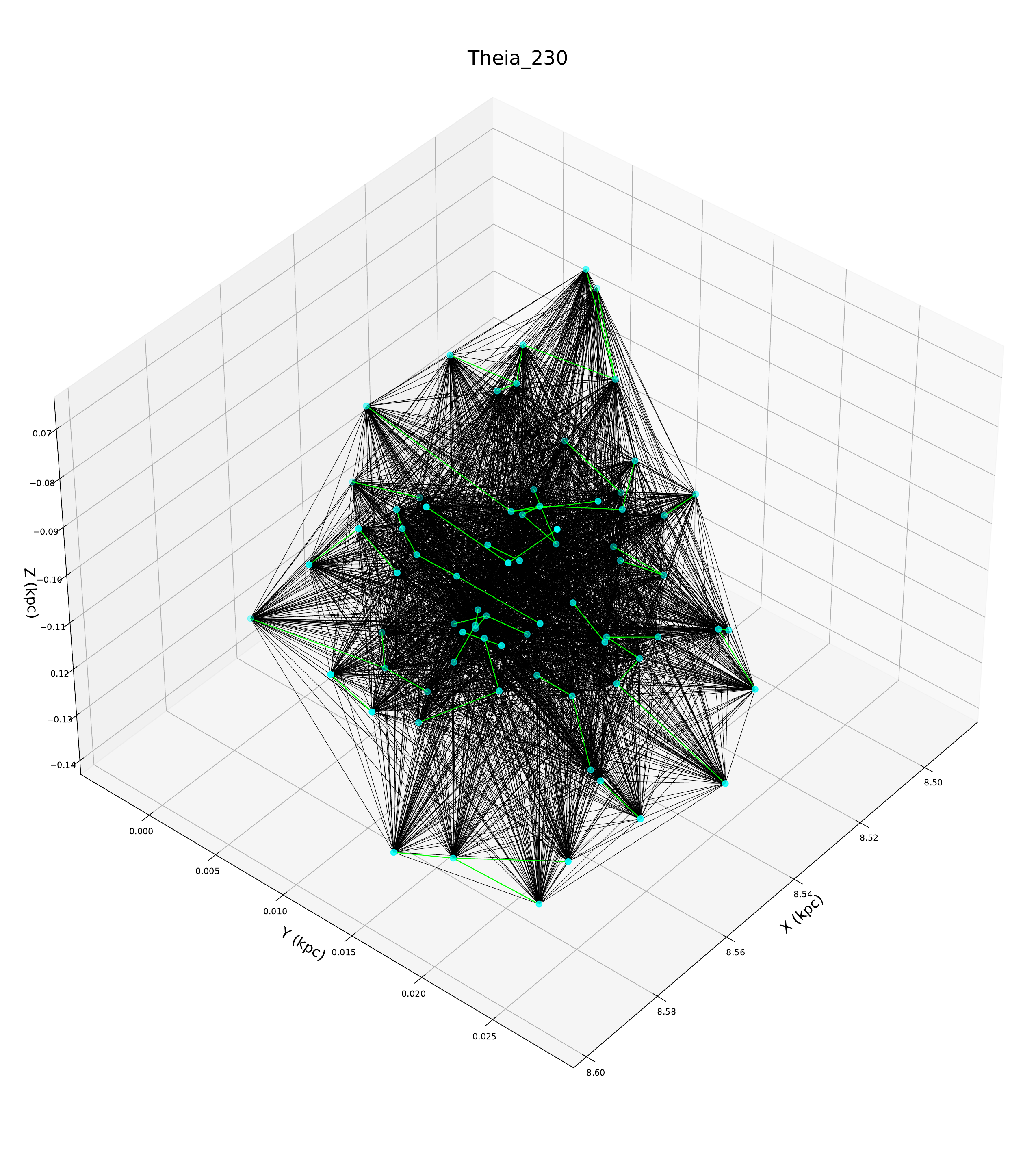}}
  \caption{Interdistances $d_{ij}$ (black colored edges) and closest interdistances ${d_{c_ij}}$ (lime colored edges) between member stars (nodes) of a "typical" young open cluster in our sample represented in 3D.}
  \label{fig:Theia_230_Di_IQR_3D}
\end{figure*}

\begin{figure*}
  \resizebox{\hsize}{!}{\includegraphics{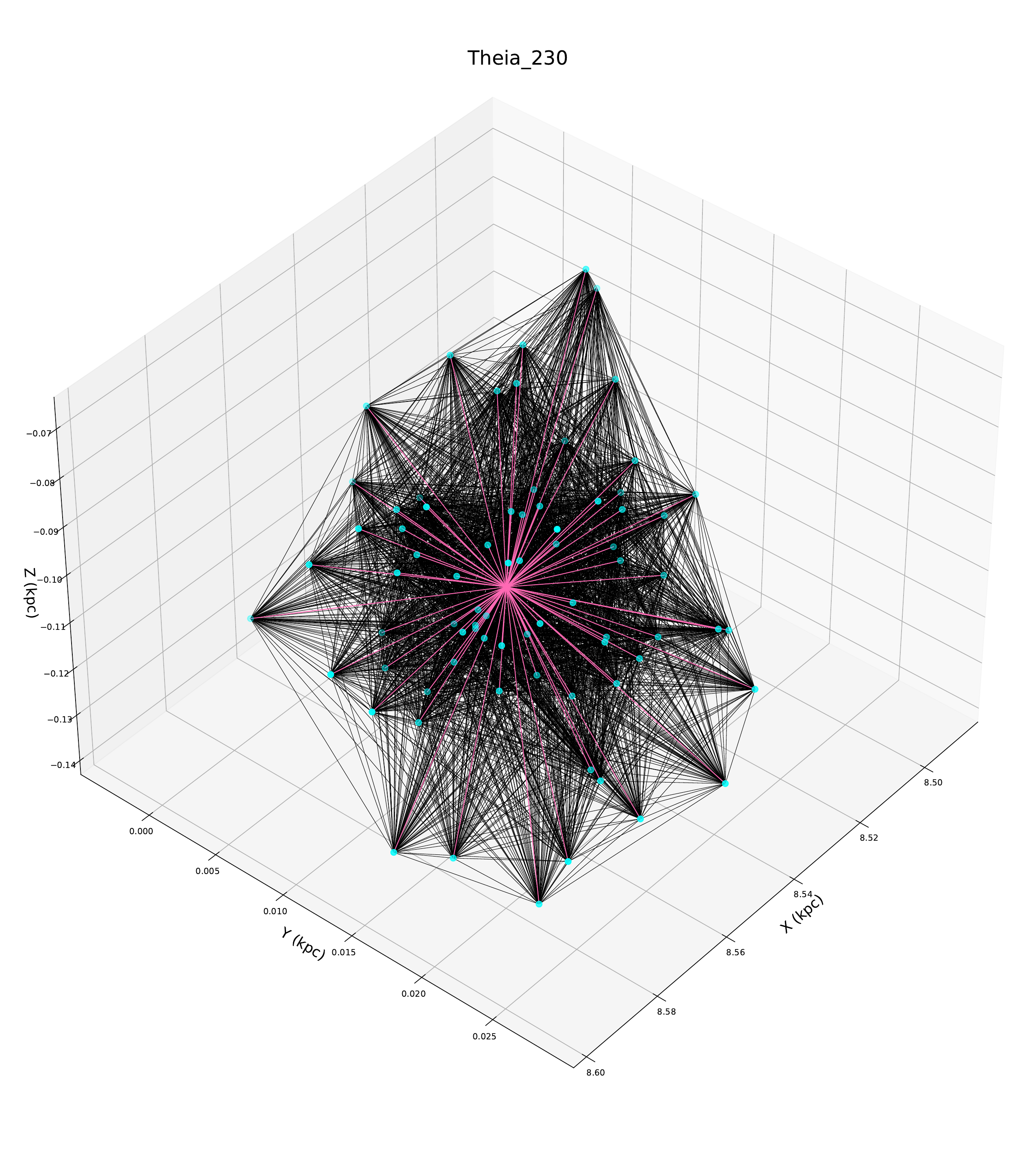}}
  \caption{Interdistances $d_{ij}$ (black colored edges) and central interdistances ${d_{cc_i}}$ (pink colored edges) between member stars (nodes) of a "typical" young open cluster in our sample represented in 3D.}
  \label{fig:Theia_230_Dc_IQR_3D}
\end{figure*}

\subsection{The median weighted central interdistance}
We introduce the concept of "median weighted central interdistance", denoted as $\bar{D_{cc}}$, which means the median distance of each member star within a cluster to the calculated weighted center of the cluster (Eq. \ref{eq:MedianDcentral}). This approach is more robust to asymmetric structures and outliers. We use G$_{mag}$ as a proxy for luminosity, considering that the member stars of a cluster are at similar distances from us. Mathematically, for a cluster with n stars and Cartesian coordinates ($x_{i},y_{i},z_{i}$), the weighted median central interdistance is computed as:

\begin{equation}
\label{eq:MedianDcentral}
\tilde{D_{\text{cc}}} = \text{med} \left( d_{\text{cc}_i} \right) = \text{med} \sqrt{(x_i - x_{\text{wc}})^2 + (y_i - y_{\text{wc}})^2 + (z_i - z_{\text{wc}})^2}
\end{equation}

where $d_{cc_i}$ is the distance from each star to the calculated centre of the cluster. The weighted centre of the cluster ($x_{wc},y_{wc},z_{wc}$) is calculated using the Cartesian coordinates weighted by a measure of magnitude (\(G_{mag}^{2/7}\)) of each star, reflecting the mass-luminosity relation \( L \propto M^{3.5} \) in the main sequence \citep{Salaris2005essp.book.....S}. We can adopt this approximation since we are considering mainly young clusters, with few or none evolved giant stars.  Mathematically, for a cluster with \(n\) stars and Cartesian coordinates \((x_i, y_i, z_i)\), the weighted centre is computed as:

\begin{equation}
\label{eq:WeightedCenter}
(x_{\text{wc}}, y_{\text{wc}}, z_{\text{wc}}) = \left( \frac{\sum_{i=1}^{n} w_i x_i}{\sum_{i=1}^{n} w_i}, \frac{\sum_{i=1}^{n} w_i y_i}{\sum_{i=1}^{n} w_i}, \frac{\sum_{i=1}^{n} w_i z_i}{\sum_{i=1}^{n} w_i} \right)
\end{equation}

where \(w_i\) is the weight of each star derived from its mean G-band magnitude:

\begin{equation}
\label{eq:Weights}
w_i = G_{mag}^{2/7}
\end{equation}

We implemented the above metrics with the {\sc numpy} library \citep{harris2020array} and {\sc scipy} \citep{SciPy20} algorithms, using the Euclidean norm to calculate the spatial distances between member stars in a three-dimensional space. We also utilized the {\sc itertools} library available in Python \citep{van1995python} to generate unique combinations of star pairs within each cluster. In Figs. \ref{fig:Theia_230_Di_IQR_3D}, \ref{fig:Theia_230_Dc_IQR_3D}, and \ref{fig:8_young_clusters_3D_with_age_Dc}, \ref{fig:8_young_clusters_3D_with_age_Dcc}, \ref{fig:8_old_clusters_3D_with_age_Dc}, \ref{fig:8_old_clusters_3D_with_age_Dcc} in the Appendix we represent the different interdistances for typical young and old clusters in our final sample represented in 3D. 

\section{The limits of {\it Gaia} {\sc dr3} in resolving star cluster structures}
\label{sec:distance}

Although {\it Gaia} has brought enormous progress in the discovery and study of star clusters, we must consider that measuring the internal spatial properties of cluster members is very sensitive, and depends enormously on the precision with which we measure distances, i.e. parallax. 
To understand what is the limit in distance at which we can correctly interpret the structure of clusters, we introduced a parameter that gives a quantitative measurement of the shape of clusters. We expect, in fact, that within the tidal radius, the clusters will not experience significant deformations. If we were to observe them, these would probably be produced by inaccurate distances for members of the same cluster. In particular, this is the case if the deformations occur mainly along the line of sight. 

\begin{figure}
  \resizebox{\hsize}{!}{\includegraphics{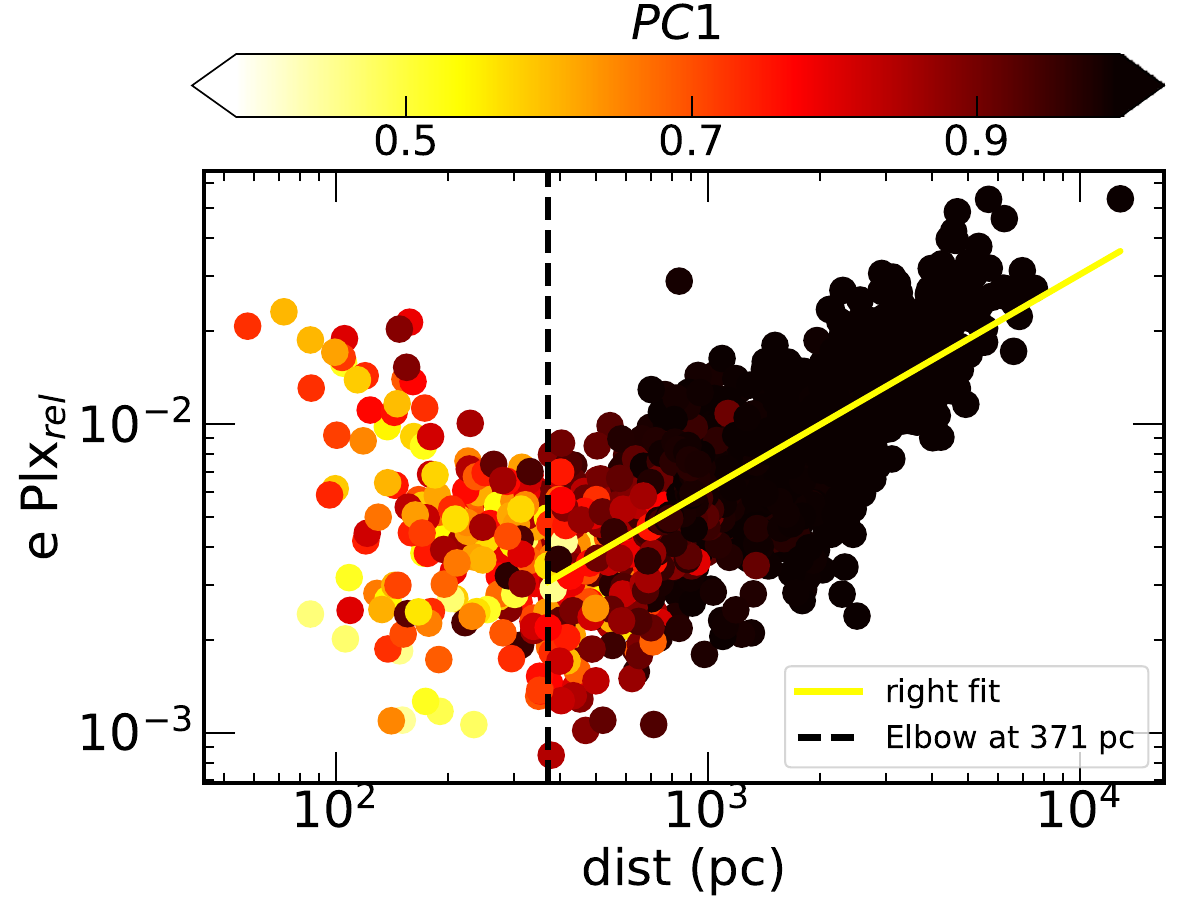}}
  \caption{Distance vs relative parallax errors for our sample of $\sim$1,500 clusters color-coded by PC1 values. The black dashed vertical line represents the slope change location.}
\label{fig:dist50_vs_e_Plx_rel}
\end{figure}

To provide a quantitative estimate of the cluster shape, we employed Principal Component Analysis (PCA), implemented using the {\sc scikit-learn} package \citep{scikit-learn11}, to analyse the distribution of stars within each cluster. The PCA method allows us to identify the principal axes along which the distribution of stars varies the most, providing insights into the shape characteristics of the clusters. By examining the variance ratios of the principal components, given by:

\begin{equation}
\label{eq:pca}
 \text{Explained Variance Ratio}\;(\text{PC}_i) = \frac{\lambda_i}{\sum_{j=1}^{N} \lambda_j}
\end{equation}

where $\lambda_{i}$ represents the eigenvalue associated with the $i$-th principal component (PC), and $N$ is the total number of principal components, we can quantify to which extent each axis contributes to the overall shape variability of the clusters. In our sample, clusters with low variance ratios along the first principal component (PC1) tend to exhibit more spherical or isotropic shapes, while those with higher ratios have elongated or anisotropic shapes. This parameter provides a quantitative way to automatically measure the shape of a cluster and classify it as compact or elongated.
In Fig.~\ref{fig:dist50_vs_e_Plx_rel} we show the relative parallax error as a function of the distance to the cluster. Excluding a small number (26 clusters) of nearby clusters for which the relative error in parallax is higher, we notice an increase in the relative error with distance, leading, as can be seen in the figure, to an apparently elongated shape of the clusters. 
We investigated the reason for higher relative errors in these 26 very nearby clusters. We noted that they contain many cool stars (T$_{\rm eff}\sim$3000 K), for which the colour correction may not be accurate \citep[see, e.g.][]{Lindegren2021A&A...649A...4L}, and also have a larger number of members with   the Renormalised Unit Weight Error {\sc RUWE} $>$ 1.4 (12\% vs. 3\% of the other clusters), indicating that  the source is non-single or otherwise problematic for the astrometric solution.

\begin{figure}
  \resizebox{\hsize}{!}{\includegraphics{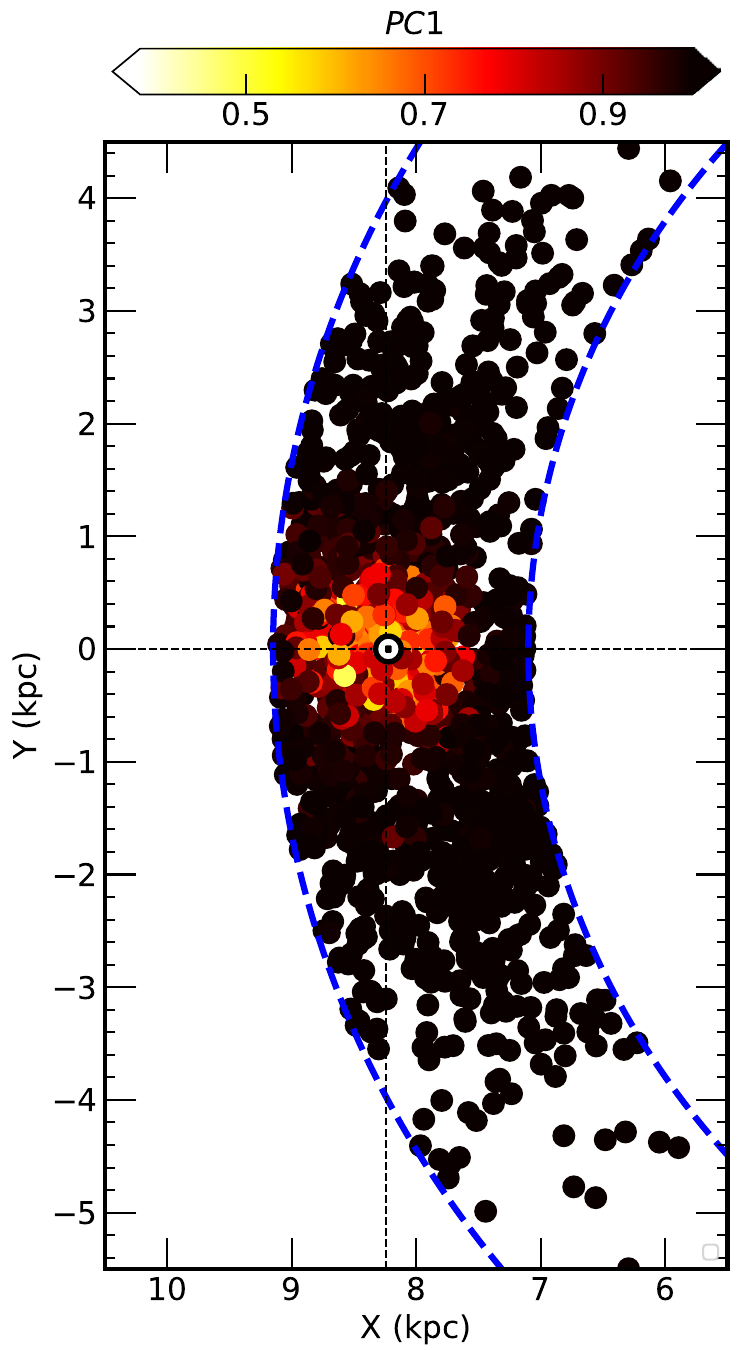}}
  \caption{Our sample of $\sim$1,500 open clusters in Galactocentric galactic coordinates X and Y, color-coded by their PC1 value. The dashed blue lines mark the region at galactocentric distance 7-9 kpc.}
\label{fig:1551_XY}
\end{figure}

\begin{figure}
  \resizebox{\hsize}{!}{\includegraphics{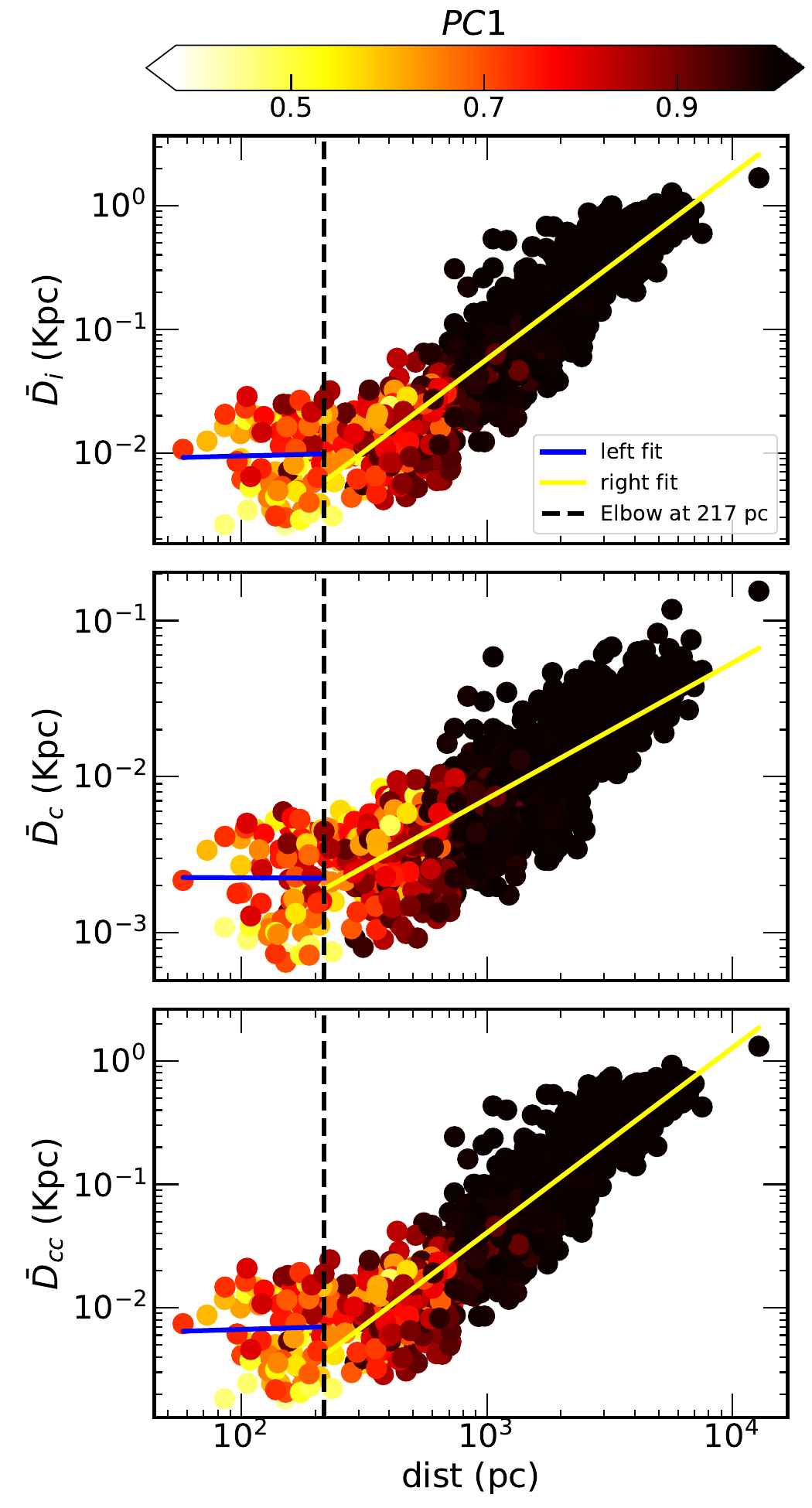}}
  \caption{$\bar{D_{i}}$, $\bar{D_{c}}$ and $\bar{D_{cc}}$ (kpc) vs distance for our sample of $\sim$1,500 open clusters color-coded by their PC1 value. The black dashed vertical line represents the slope change location. The solid yellow and blue lines are a linear fit to the data at both intervals of slope change point.}
\label{fig:distance_vs_3_interdistances_IQR_solar_Q1-Q3_vs_PC1}
\end{figure}

In Figs.~\ref{fig:1551_XY} and \ref{fig:distance_vs_3_interdistances_IQR_solar_Q1-Q3_vs_PC1}, we show the location of the clusters in the Galactic Plane and from the Sun, respectively,   highlighting the variation in their PC1. From these figures, we notice that the greater the distance from the Sun, the greater the distortion of the clusters, mainly along the line-of-sight. 
Fig.\ref{fig:distance_vs_3_interdistances_IQR_solar_Q1-Q3_vs_PC1} shows a clear change of slope in the relationship between the interdistance and distance. We calculated this change in slope using the Kneedle algorithm \citep{Satopaa11} as in \citet{magrini23}, finding that it occurs at a distance of 217 pc. Clusters more far away than this distance have a drastic increase in their interdistances. Most of them seem to expand and align along the line-of-sight.  
This phenomenon arises due to the considerably larger uncertainty associated with parallaxes compared to the uncertainties on the  positions on the plane of the sky. Consequently, the location of cluster members appears significantly more extended in this particular direction.  It is important to emphasize that this is an observational bias and does not reflect the true distribution of stars within a cluster \citep[see, e.g.][for the globular cluster population]{Vasiliev2021MNRAS.505.5978V}. 
Since this issue becomes more prominent at distances larger than 220 pc  and exacerbates with increasing distance, to avoid it, we restrict our final sample, here after ''benchmark sample'', to distances less than or equal to about 220 parsec.  Our benchmark sample is composed of 81 clusters.

\section{The time evolution of cluster interdistances}
\label{sec:interditances_vs_agewe}

In this Section, we discuss the correlations between the various definitions of interdistance and the ages of star clusters, expressed in logarithmic form, in our benchmark sample (81 clusters, within $\sim$220 pc). Our aim is to study how cluster properties evolve over time.

\begin{figure}
  \resizebox{\hsize}{!}{\includegraphics{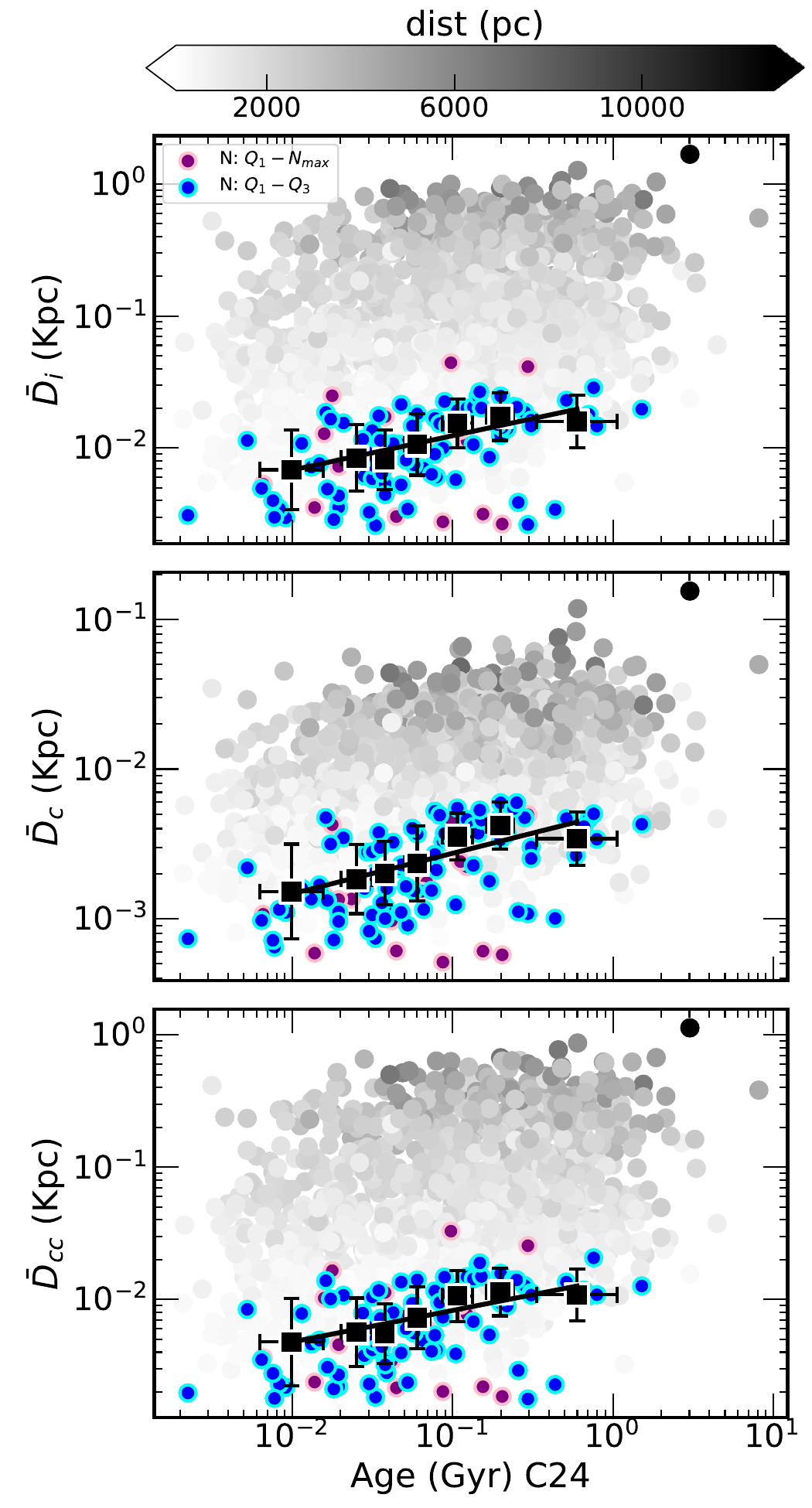}}
  \caption{$\bar{D_{i}}$, $\bar{D_{c}}$ and $\bar{D_{cc}}$ (kpc) vs Age (Gyr) on a logarithmic scale for a subsample of 81 open clusters (blue) at a distance < 217 pc. The clusters are also shown in 7 equally distributed bins (black colour). In the background (light gray), the our sample of $\sim$1,500 clusters in the solar region. The black lines show the linear fit on a logarithmic scale to the clusters' data. The most populated open clusters at a distance < 217 pc are shown in purple.}
  \label{fig:Age_Cavallo_vs_Di_Dc_Dcc}
\end{figure}
 
In Fig.\ref{fig:Age_Cavallo_vs_Di_Dc_Dcc}, we show  the interdistances $\bar{D_{i}}$, $\bar{D_{c}}$ and $\bar{D_{cc}}$ as a function of the age (Gyr) on a logarithmic scale. By employing a logarithmic scale, we can effectively capture the full range of cluster ages, from the relatively frequent young clusters to the less common older ones, thus providing a comprehensive perspective on the evolutionary timeline of open clusters. 
In these figures we show the complete sample, but we focus for analysis and we perform the fit only on the benchmark sample.

\begin{figure}
  \resizebox{\hsize}{!}{\includegraphics{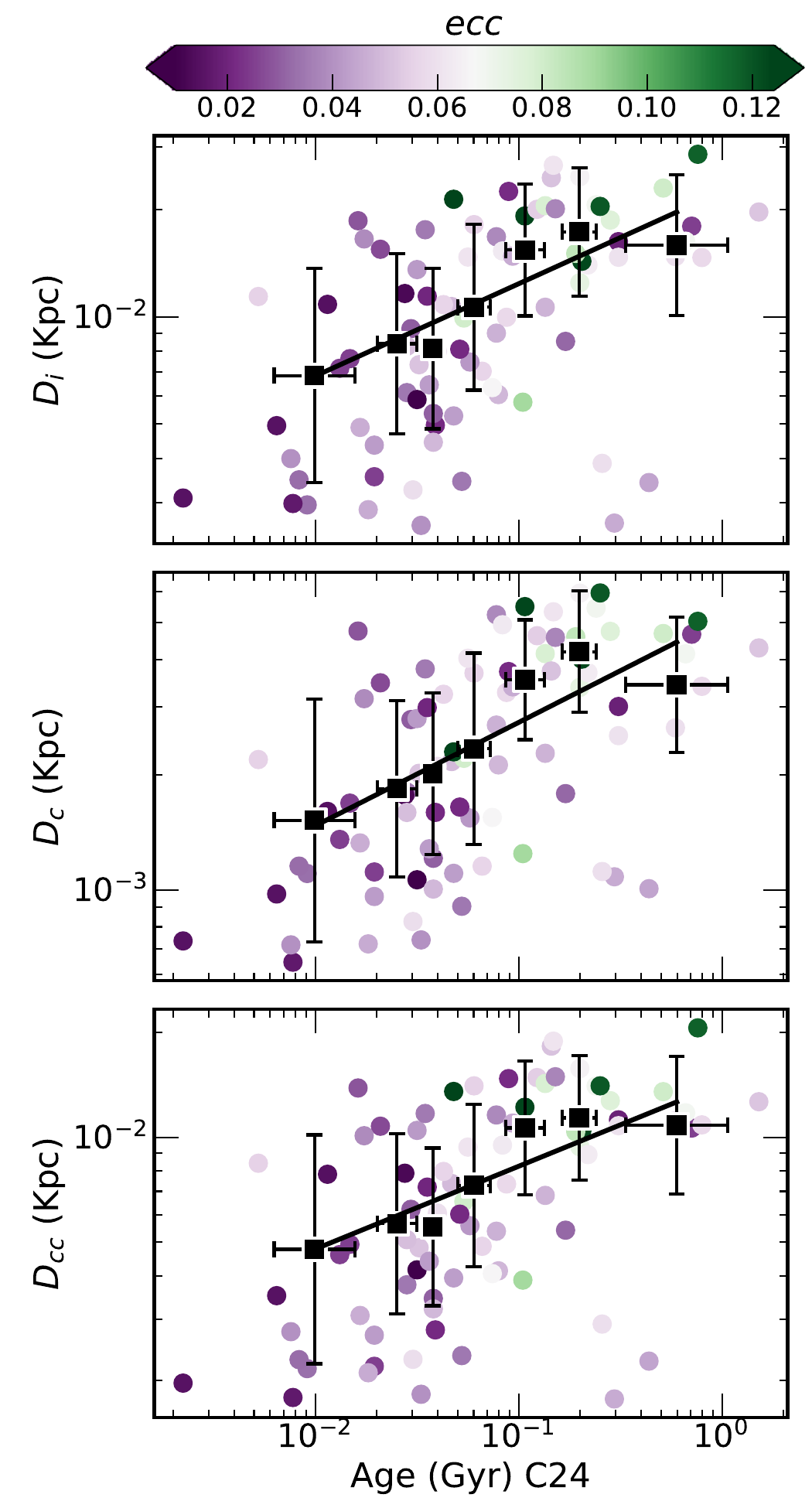}}
  \caption{$\bar{D_{i}}$, $\bar{D_{c}}$ and $\bar{D_{cc}}$ (kpc) vs Age (Gyr) on a logarithmic scale for the subsample of 81 open clusters at a distance < 217 pc colour coded by $ecc$. The clusters are also shown in 7 equally distributed bins (black colour).}
  \label{fig:112_Age_Cavallo_vs_Di_Dc_Dcc_by_ecc}
\end{figure}

\begin{figure}
  \resizebox{\hsize}{!}{\includegraphics{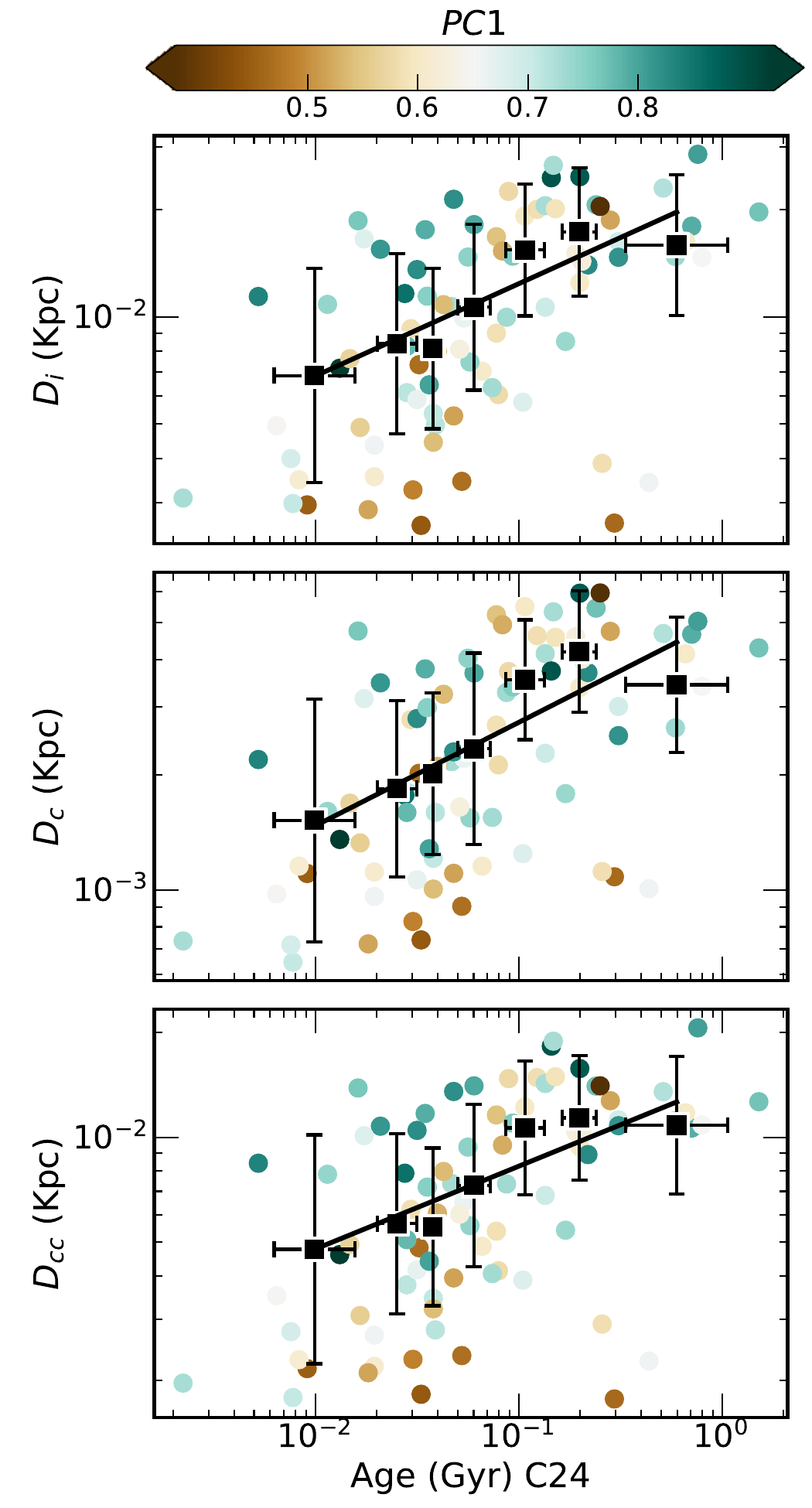}}
  \caption{$\bar{D_{i}}$, $\bar{D_{c}}$ and $\bar{D_{cc}}$ (kpc) vs Age (Gyr) on a logarithmic scale for the subsample of 81 open clusters at a distance < 217 pc colour coded by $PC1$. The clusters are also shown in 7 equally distributed bins (black colour).}
  \label{fig:112_Age_Cavallo_vs_Di_Dc_Dcc_by_PC1}
\end{figure}

As can be seen, there is a clear correlation of the three interdistances with age. We divided the 81 open clusters into 7 equally distributed bins representing 16 open clusters each. We used \texttt{TheilSenRegressor} implemented using the {\sc scikit-learn} package \citep{scikit-learn11} to perform robust line-fit. The Kendall-Theil-Sen regression \citep{Theil1950, sen1968} has several advantages over Ordinary Least Squares regression, being less sensitive to outliers. This method computes the median of all slopes between pairs of points, making it less affected by extreme values. The coefficients of the regression to the data of the restricted sample are defined by the equation \ref{eq:log_fit_Di} and are shown in Tab. \ref{tab:regressions_all}.

\begin{equation}
\label{eq:log_fit_Di}
\log_{10}(\bar{D_{x}}) = m \cdot \log_{10}(age) + c
\end{equation}

\begin{table}
\caption{Linear regression coefficients (slope and y-intercept) of the Age vs interdistances relation on a logarithmic scale for the restricted  sample obtained using the Kendall-Theil-Sen robust line-fit method, as well as Pearson and Spearman correlation coefficients and their p-values.}
\label{tab:regressions_all}
\scalebox{0.95}{
\begin{tabular}{lcccccc}

\hline\hline
Param. & m & c & PCC & p-value & SCC & p-value \\\hline
$\bar{D_{i}}$ & 0.259 & $-$1.65 & 0.91  & 0.004 & 0.93 & 0.002 \\
$\bar{D_{c}}$ & 0.270 & $-$2.29 & 0.89 & 0.007 & 0.89 & 0.007 \\
$\bar{D_{cc}}$ & 0.238 & $-$1.84 & 0.91 & 0.004 & 0.93 & 0.002 \\

\hline
\end{tabular}}
\end{table}

In the central panel of Fig. \ref{fig:Age_Cavallo_vs_Di_Dc_Dcc} we show the evolution of the $\bar{D_{c}}$ over time. As in the case of $\bar{D_{i}}$, we find a strong correlation, with PCC and SCC $\simeq$ 0.9 and low p-values (see Tab. \ref{tab:regressions_all}). Finally, in bottom panel of Fig. \ref{fig:Age_Cavallo_vs_Di_Dc_Dcc} we show the evolution of the $\bar{D_{cc}}$ over time. As in the two previous cases, we find a strong correlation, with PCC and SCC $\simeq$ 0.9 and low p-values (see Tab. \ref{tab:regressions_all}). The behaviour is very similar to that of the $\bar{D_{i}}$ case, although the y-intercept is slightly lower.
In all three cases, we have a clear trend of increasing interdistance with time, which is surely linked to the internal evolution  \citep[see, e.g.][]{Miret2023}, together with external effect of perturbations \citep[see, e.g.][]{Viscasillas23}. 

From Fig.~\ref{fig:Age_Cavallo_vs_Di_Dc_Dcc} we deduce that given a cluster with a typical number of members, its interdistances, both between members and with respect to the centre, correlate very well with the age (from isochrones) of the cluster. However, we have to notice that our age range extends down to 1 Gyr, and there appears a tendency for the relationship to flatten out for the older ages. This may be related to the slowing down of expansion and reaching a state of equilibrium for the few surviving older clusters \citep[see, also][]{Tarricq2022}. We also evaluated the effect of  including the most populated clusters (without restriction on the maximum number of members) and in the same range of distances as the benchmark sample. As we can see in the Fig.~\ref{fig:Age_Cavallo_vs_Di_Dc_Dcc}, the most massive clusters are located around the relationship drawn by the typical clusters. Some of them, however, maintain smaller interdistances as time passes, as if their larger masses better preserved them from expansion. 

Thus, once we have calibrated a relationship between age and interdistance, we can estimate the age of clusters of which we only know the spatial arrangement of the members. This relationship is obviously valid in the range in which we can calibrate it, i.e. up to about 1 Gyr, and in the distance range of our benchmark sample. 
The precision of this method depends not only on the quality of the calibration, but also on whatever may influence the dynamic evolution of a cluster, starting with the initial conditions up to the perturbations that may occur during its lifetime. 
From Fig.~\ref{fig:Age_Cavallo_vs_Di_Dc_Dcc}, we also notice a similar behaviour in the larger sample of clusters (the shaded grey area in the plot). In the large sample, the relation obtained with the benchmark sample is artificially shifted upwards by parallax uncertainties that increase with distance. 
However, a constant offset is maintained independent of the age of the cluster, which in principle would allow us to define similar relationships in which the distance is included as an additional parameter. 
We do not include these relationships, because they are based not on physical parameters, but on the dependence of parallax error on the distance. 

\subsection{Effects related to the selection criterion for members}
The approach of selecting stars with a high probability of membership is driven by the need to avoid contaminants. However, this means considering about 44\% of all possible members. We therefore tested the effect of relaxing the constraint on the probability with P > 0.7. We observed that the members with lower P are preferentially located in the outer regions of the clusters, and thus tend to increase the interdistances, particularly those from the centre.
Following the same approach we used for the selection with P > 0.9, we computed the relationships between cluster ages and interdistances. The relationships maintain the same shape, with a larger scatter -probably due to the increased number of contaminants- and an upward shift due to the inclusion of more outlying members, which increases, on average, the interdistances. The general conclusions of our work are, therefore, not affected by this choice. 

\subsection{External perturbative effects}
To estimate the external perturbative effects on the time evolution of interdistances, we considered two parameters: the eccentricity,  $ecc$, of the orbit and and the shape of the clusters, measured through its PC1. 
In the Figs. $\ref{fig:112_Age_Cavallo_vs_Di_Dc_Dcc_by_ecc}$ and $\ref{fig:112_Age_Cavallo_vs_Di_Dc_Dcc_by_PC1}$, we present the  interdistances vs age, highlighting the dependence on $ecc$ and PC1.  
Departure from circular orbits increases with increasing stellar population age, thus the $ecc$ increases  with age, as already noticed in \citet{Viscasillas23} and discussed in the next section of the present work. 
From Figs. $\ref{fig:112_Age_Cavallo_vs_Di_Dc_Dcc_by_ecc}$ we notice that older clusters,  with larger interdistances,  are also those with higher $ecc$. 
However,  in our benchmark sample orbits are always very close to circularity ($ecc$ hardly ever exceeds 0.1) because the sample reaches a maximum of 1 Gyr in age. 
Moreover, at a given age, there is not a clear separation between clusters on circular orbits or clusters with slightly more eccentric orbit. 
In our benchmark sample,  the dominant effect in the expansion of young clusters seems to be the intrinsic one as opposed to that due to the external environment that causes perturbations on the orbit. 

In Fig.~$\ref{fig:112_Age_Cavallo_vs_Di_Dc_Dcc_by_PC1}$ we analyse the effect of  PC1, i.e. preferential direction in the cluster shape. In general, there is no large effect of PC1 on the relationship between age and interdistances: more spherical or slightly elongated clusters seem to maintain the same evolution in the interdistances between their members. However, we  note that the some of the few clusters with PC1 < 0.5, i.e. the more spherical ones, are usually those with lower interdistances, located well below the correlation. These few clusters do not seem to show an evolution with time of the interdistances. They may be those born as denser and  bound systems. Comparing with Fig.~\ref{fig:Age_Cavallo_vs_Di_Dc_Dcc}, the sequence of round compact clusters, below the correlation line, is also populated by several massive clusters. This strengthens the result shown in \citet{Viscasillas23} that the most massive and compact clusters are those most likely to survive over time. 

So, in conclusion, our sample shows that internal cluster expansion effects, which are related also to the mass and density of clusters,  are the dominant ones in shaping the relationship between interdistances and age.

\section{Dependence of orbital parameters, $ecc$ and Z$_{max}$, on cluster age}
\label{sec:orbits}

In this section, we aim to analyze how the orbital  parameters  can be related to the cluster properties, in particular their ages. 
\citet{Viscasillas23} found that the orbits of older clusters are more perturbed with respect to the younger ones. The old clusters are usually found at higher heights above the plane (|Z$_{\rm max}$| and with orbits of larger $ecc$. 
As in this section  we do not use the information on the internal structure of the clusters, we can verify the existence of such correlations between age and orbital parameters for the full sample of 1,500 clusters in the solar neighbourhood.  
We calculated a linear fit between age and $ecc$, and age and |Z$_{\rm max}$| using {\sc statsmodels} routines \citep{seabold2010statsmodels}. 
Specifically, we obtained:
\begin{equation}
    ecc = 0.049\pm 0.007 + age \times 0.140\pm0.004
\label{eq7}
\end{equation} and 
\begin{equation}
    |Z_{\rm max}| = 0.125\pm 0.005 + age \times 0.088\pm0.003
\label{eq8}
\end{equation}
in which age is expressed in Gyr and |Z$_{\rm max}$| in kpc. 
Both relationships are very similar to that obtained in \citet{Viscasillas23} with a smaller number of clusters, also here reported:
\begin{equation}
ecc = 0.030\pm0.006 + age \times 0.049\pm0.015 
 \label{eq9}
\end{equation} 
and 
\begin{equation}
|Z_{\rm max}| = 0.124\pm0.022 + age \times 0.051\pm0.049  
 \label{eq10}
\end{equation}

Furthermore, when we calculate the same relations for the benchmark sample (dist < 217 pc), we obtain the following:

\begin{equation}
    ecc = 0.030\pm 0.012 + age \times 0.046\pm0.003
     \label{eq11}
\end{equation} and 
\begin{equation}
    |Z_{\rm max}| = 0.036\pm 0.023 + age \times 0.077\pm0.006.
    \label{eq12}
\end{equation}

\begin{figure}
\resizebox{\hsize}{!}{\includegraphics{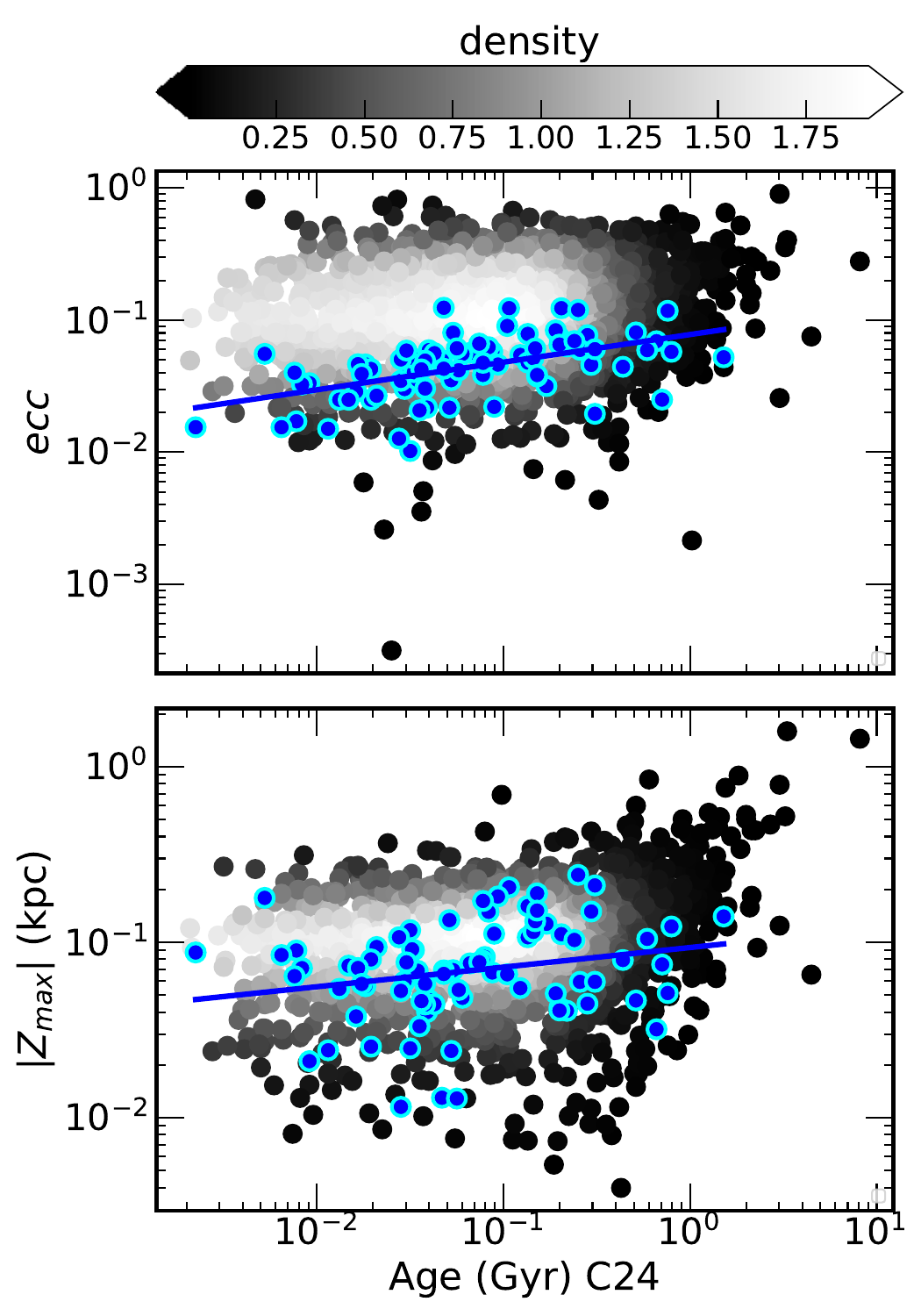}}
\caption{Age C24 (Gyr) vs $ecc$ and $|Z_{max}|$ (kpc) for our sample of $\sim$1,500 clusters colored using a point density function. The blue points correspond to the benchmark  sample located at distances < 217 pc, and the solid blue line with its corresponding linear fit.}
\label{fig:Age_Cavallo_vs_e_zmax}
\end{figure}

Thus, both our samples, the  largest one and the benchmark one, confirm the previous results \citep[cf.][our eqs. \ref{eq9} and  \ref{eq10}]{Viscasillas23}, as indicated by the equations \ref{eq7}, \ref{eq8}, \ref{eq11} and \ref{eq12}.
Therefore, as time passes, the orbits of the clusters become more perturbed, spotting them from a purely circular orbit that lies on the Galactic plane. 
In Fig. $\ref{fig:Age_Cavallo_vs_e_zmax}$,  we show $ecc$ and $|Z_{max}|$ as a function of cluster age. To visualize the relationships, we utilised a point density function (\texttt{gaussian\_kde}) implemented using {\sc scipy.stats} \citep{SciPy20}. For a  better visualisation, the figure is shown on a logarithmic scale. As we can see, most of the clusters remain in circular orbits during the first 500 million years, which corresponds to the epoch in which we have the largest number of clusters. From that epoch, clusters begin to be disrupted, as the cluster density drops,  and those that survive experience a sudden change in the orbital parameters, with a growth in the orbit $ecc$ and $|Z_{max}|$.

\section{Summary and conclusions}
\label{sec:conclusions}

In this work, we provide a comprehensive investigation into the evolution of the spatial distribution of stars in open clusters. We investigated the relation between dynamical properties and the cluster age. We used a sample of cluster members from the database of  \citet{Hunt23}. We  applied several selection criteria to obtain a reliable sample of cluster members, belonging to approximately 1,500 open clusters, These clusters are located around the Solar circle, with Galactocentric distances between 7 and 9 kpc.
Our analysis focused on three quantities: the Mean Interdistance ($\bar{D_{i}}$), the Mean Closest Distance ($\bar{D_{c}}$), and the Median Weighted Central Distance ($\bar{D_{ cc}}$), which are different ways to measure the distances between cluster members.  


We measured the shapes of clusters through a Principal Component Analysis (PCA). We found  that clusters with isotropic shapes tend to have lower interdistances, while clusters whose members follow a preferred direction show higher interdistances. 
However, for distances larger than about 220 pc, {\it Gaia} data are not able to resolve the internal cluster structures. For those clusters, our PC1 analysis is only able to capture the artificial elongation along the line-of-sight. 

Thus, since the correlation between the ages and interdistances for distant clusters is partly influenced by parallax errors, we reduce our cluster sample within a maximum distance of approximately 220 parsecs to mitigate observational biases and ensure that clusters remain undisturbed along the line-of-sight. With our benchmark cluster sample, limited in distance, we have found a clear relationship between interdistances and cluster age. This relationship appears to be valid for clusters located even farther away, although the current constraints on parallax precision hinder our ability to distinguish between observational artefacts and real phenomena. 
We also investigated the effect of orbit perturbation and cluster shape on the relationship between age and interdistances. These quantities have a secondary effect on the internal evolution and expansion of the clusters. 
Finally, we confirmed and extended the results of \citet{Viscasillas23} on the time  evolution of eccentricity and $|Z_{max}|$ in the sample of 1,500 clusters in the solar neighbourhood.

This work shows the complex interplay between the internal evolution of clusters and the perturbative effect of the environment, which modifies both their internal structure and their orbit in the Galactic potential. 
Although perturbations modify the orbits of clusters, they do not seem to be dominant in driving the expansion with time of clusters. 
Further exploration of this correlation promises to deepen our understanding of cluster dynamics, offering potential insights into the derivation of more precise cluster dynamical ages. Comparison with simulations of evolving star cluster populations can help to explore the variation in cluster lifetimes due to specific initial conditions and characteristics  \citep[see, e.g][]{Baumgardt2007MNRAS.380.1589B,  Whitehead2013ApJ...778..118W, Kamdar2019ApJ...884..173K, Krause2020SSRv..216...64K, Farias2024MNRAS.527.6732F}.



\begin{acknowledgements}
This work made use of Astropy \citep{Astropy_2018}, Scikit-learn Machine Learning \citep{scikit-learn11}, Scipy \citep{SciPy20}, Seaborn \citep{Waskom2021}, TopCat \citep{Taylor2005}, Statsmodels \citep{seabold2010statsmodels}, Pandas \citep{pandas2020} and Matplotlib \citep{Hunter2007}. CVV and LM thanks the EU programme Erasmus+ Staff Mobility for their support. CVV and LM thanks INAF for the support (MiniGrant Checs, Large Grant EPOCH). LM acknowledge financial support under the National Recovery and Resilience Plan (NRRP), Mission 4, Component 2, Investment 1.1, Call for tender No. 104 published on 2.2.2022 by the Italian Ministry of University and Research (MUR), funded by the European Union – NextGenerationEU– Project ‘Cosmic POT’  Grant Assignment Decree No. 2022X4TM3H  by the Italian Ministry of Ministry of University and Research (MUR).
CVV and GT acknowledge funding from the Research Council of Lithuania (LMTLT, grant No. P-MIP-23-24). This work benefited from fruitful discussions during the Workshop 'Form star cluster to field population' held in Firenze (Italy), November 2023. We thank the anonymous referee for useful comments that enriched the paper considerably.

\end{acknowledgements}

\bibliographystyle{aa} 
\bibliography{revised_version}

%
%
\begin{appendix}
\section{Complementary material}
\label{Appendix}

\begin{table*}
\caption{Main properties of the final sample of 81 open clusters used in this study.}
\label{tab:OCproperties}
\scalebox{0.97}{
\begin{tabular}{lcccccccccc}

\hline\hline
Cluster & $\bar{D_{i}}$ (kpc) & $\bar{D_{c}}$ (kpc) & $\bar{D_{cc}}$ (kpc) & Age (Gyr) & R$_{gc}$ (kpc) & $e$ (kpc) &  Z$_{max}$ (kpc) & PC1 & PC2 & PC3 \\
\hline
  ADS\_16795 & 0.017 & 0.004 & 0.012 & 0.035 & 8.199 & 0.035 & 0.069 & 0.790 & 0.126 & 0.083\\
  Alessi\_9 & 0.004 & 0.001 & 0.003 & 0.257 & 7.918 & 0.060 & 0.059 & 0.585 & 0.234 & 0.181\\
  CWNU\_1010 & 0.006 & 0.002 & 0.004 & 0.079 & 8.254 & 0.049 & 0.082 & 0.580 & 0.294 & 0.126\\
  CWNU\_1015 & 0.008 & 0.002 & 0.006 & 0.040 & 8.265 & 0.059 & 0.047 & 0.531 & 0.404 & 0.065\\
  CWNU\_1018 & 0.011 & 0.002 & 0.007 & 0.135 & 8.251 & 0.048 & 0.161 & 0.700 & 0.191 & 0.109\\
\hline
\end{tabular}}
\tablefoot{Only a segment of this table is displayed here. A machine-readable version of the complete table is accessible online.}
\end{table*}

\begin{figure*}
  \resizebox{\hsize}{!}{\includegraphics{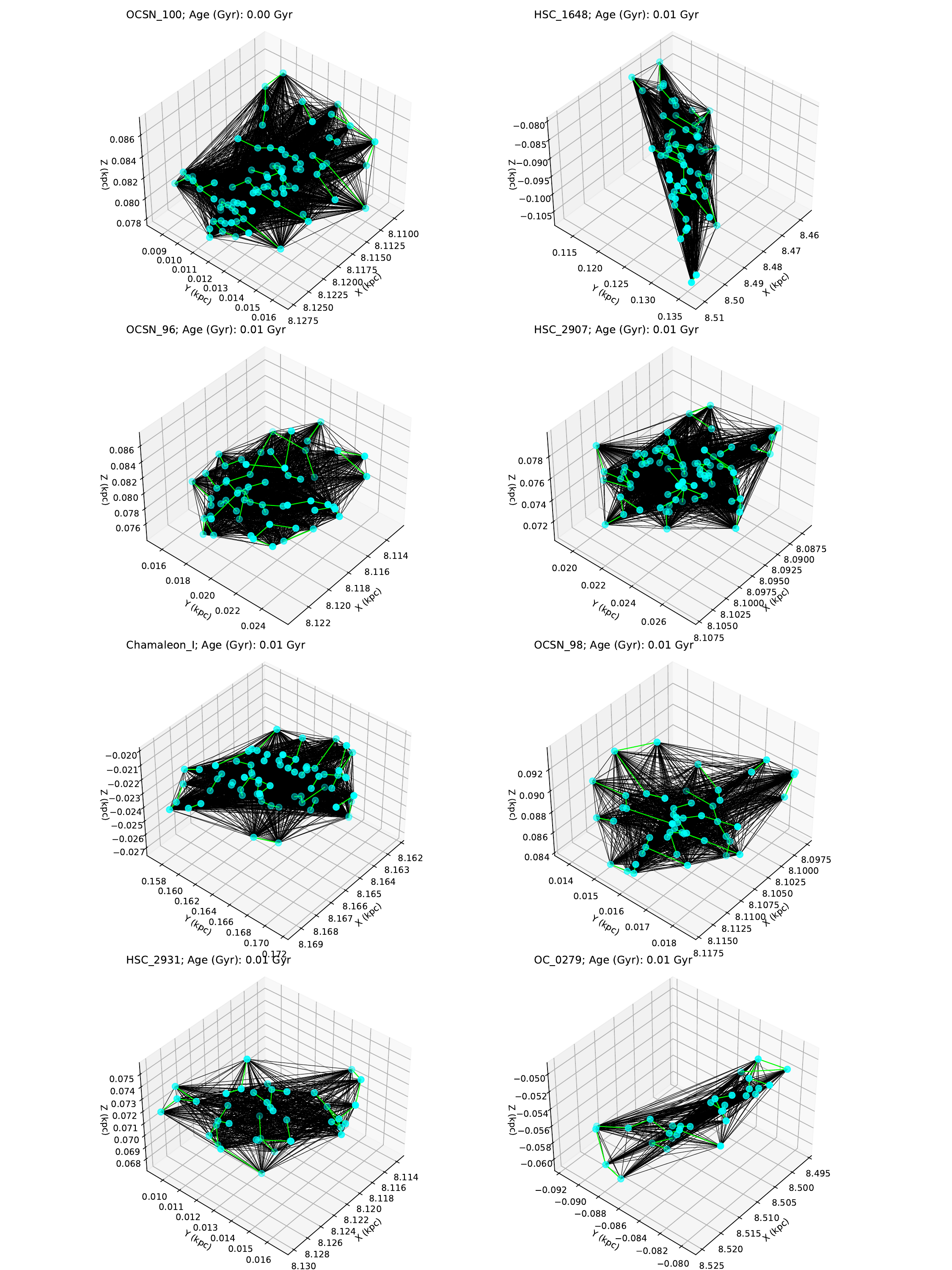}}
  \caption{Interdistances $d_{i}$ (black colored edges) and closest interdistances ${d_{c_i}}$ (lime colored edges) between member stars (nodes) of typical young open clusters of our final sample represented in 3D.}
  \label{fig:8_young_clusters_3D_with_age_Dc}
\end{figure*}

\begin{figure*}
  \resizebox{\hsize}{!}{\includegraphics{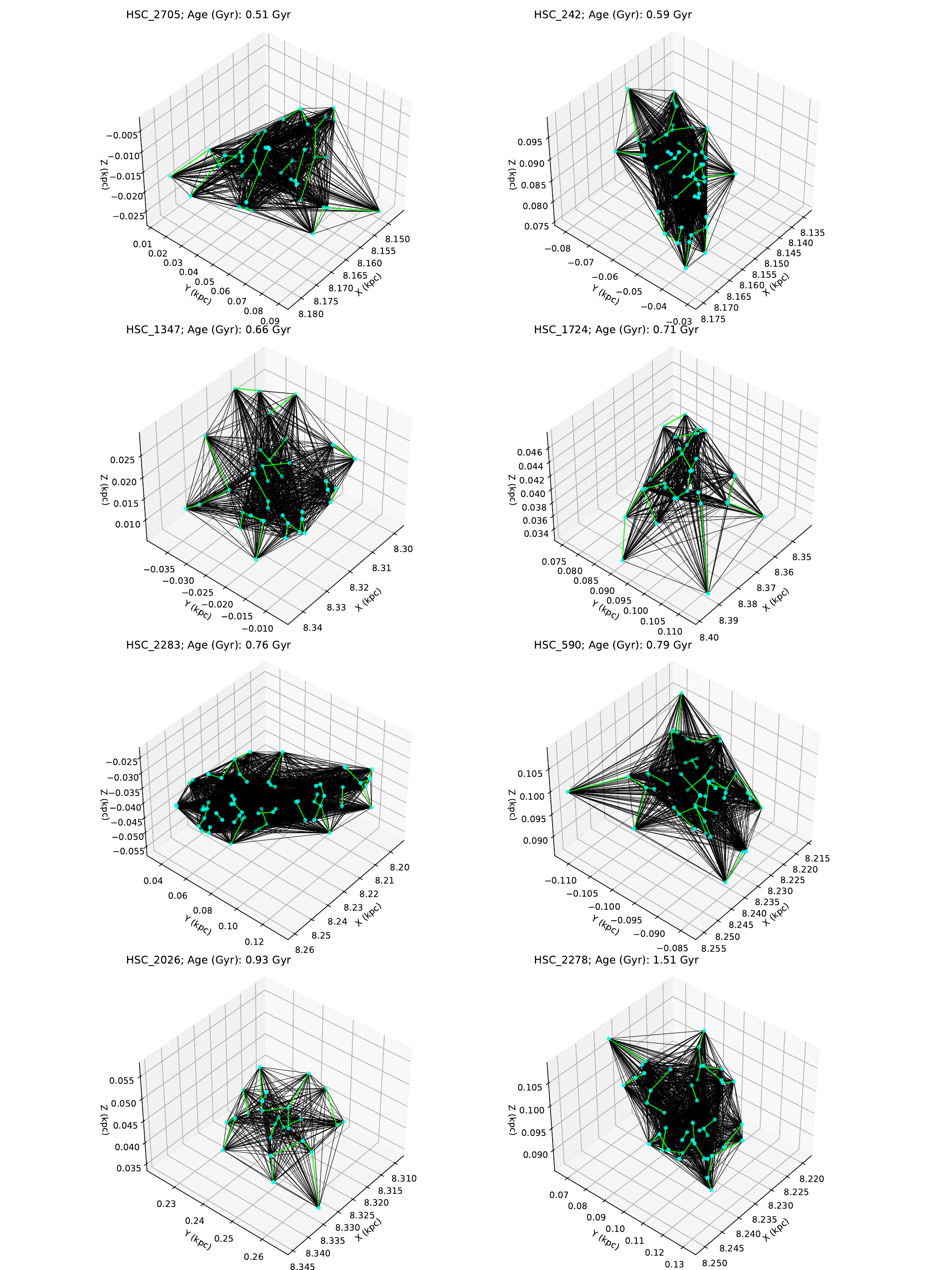}}
  \caption{Interdistances $d_{i}$ (black colored edges) and closest interdistances ${d_{c_i}}$ (lime colored edges) between member stars (nodes) for typical old open clusters of our final sample represented in 3D.}
  \label{fig:8_old_clusters_3D_with_age_Dc}
\end{figure*}

\begin{figure*}
  \resizebox{\hsize}{!}{\includegraphics{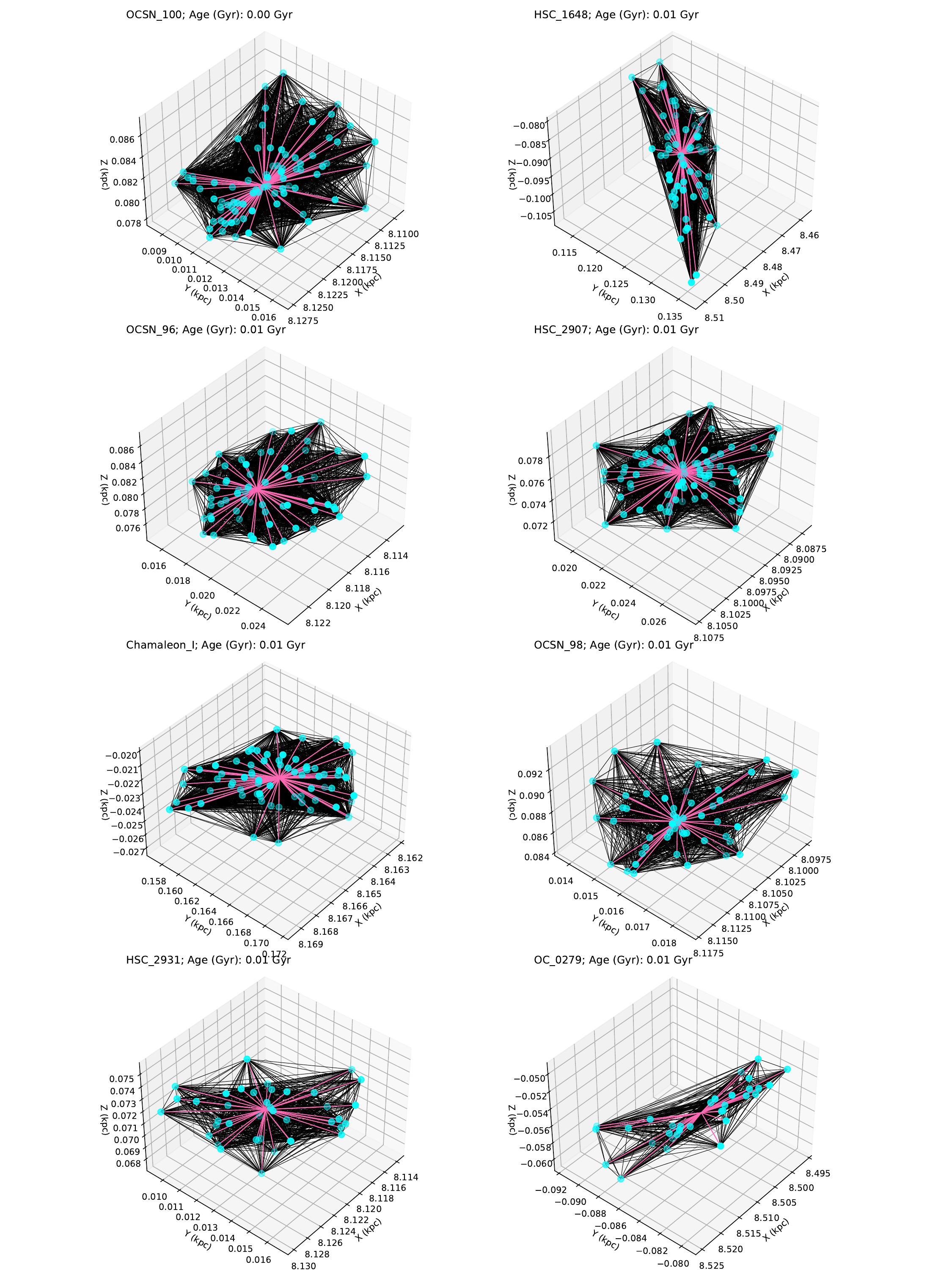}}
  \caption{Interdistances $d_{i}$ (black colored edges) and central interdistances ${d_{cc_i}}$ (pink colored edges) between member stars (nodes) of typical young open clusters of our final sample represented in 3D.}
  \label{fig:8_young_clusters_3D_with_age_Dcc}
\end{figure*}

\begin{figure*}
  \resizebox{\hsize}{!}{\includegraphics{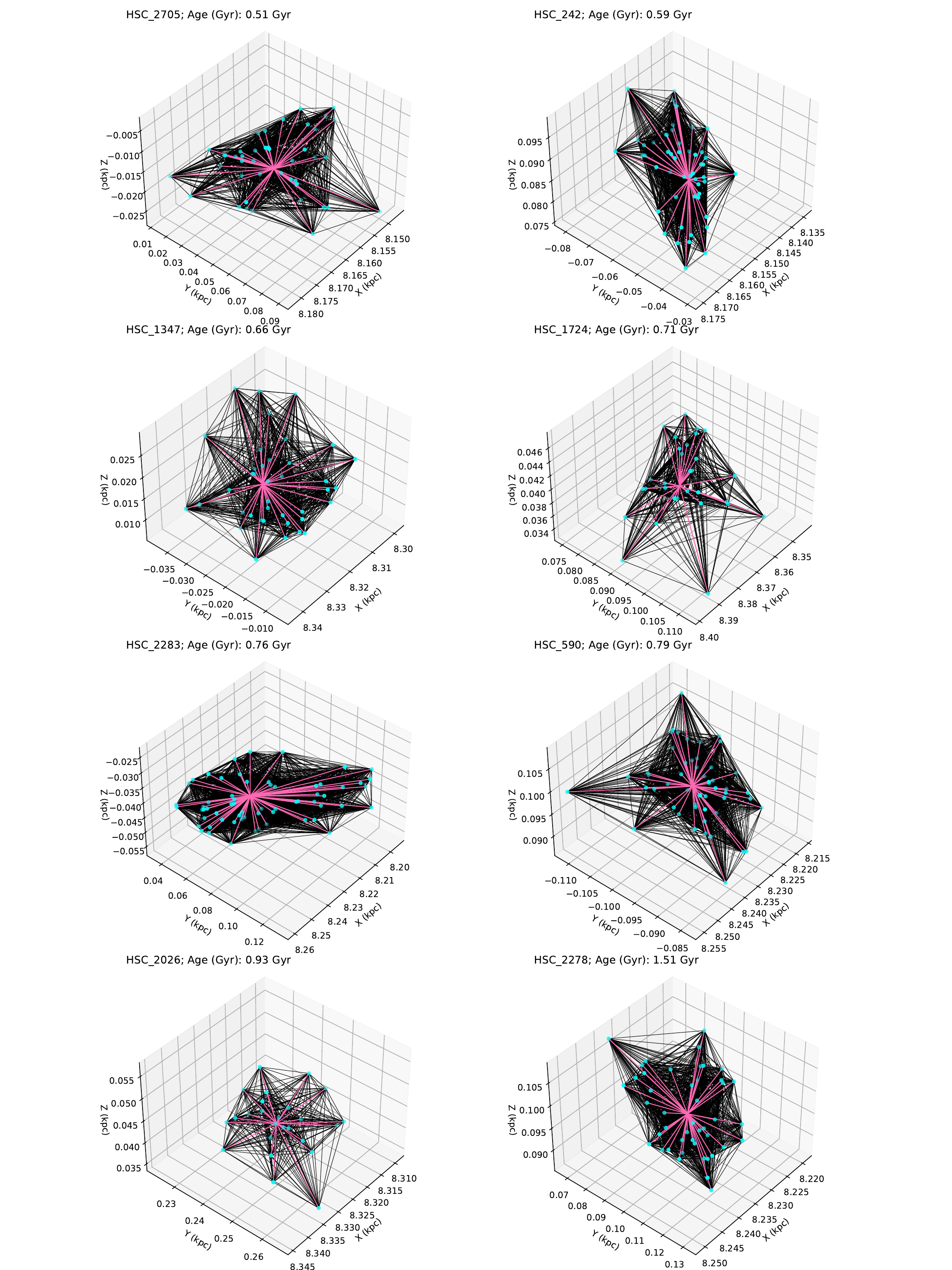}}
  \caption{Interdistances $d_{i}$ (black colored edges) and central interdistances ${d_{cc_i}}$ (pink colored edges) between member stars (nodes) of typical old open clusters of our final sample represented in 3D.}
  \label{fig:8_old_clusters_3D_with_age_Dcc}
\end{figure*}


\end{appendix}

\end{document}